# Juno: a Python-based graphical package for optical system design

**David Dierickx,**[1] **Patrick Cleeve,**[1] **Sergey Gorelick,**[2] **James C. Whisstock,**[1] **Alex de Marco,**[1, *]

[1] *Department of Biochemistry and Molecular Biology, Monash University, 3800 Clayton, Victoria (Australia)*
[2] *Monash Centre for Electron Microscopy, Monash University, 3800 Clayton, Victoria (Australia)*
*Corresponding author: alex.demarco@monash.edu*

This report introduces Juno, a modular Python package for optical design and simulation. Juno consists of a complete library that includes a graphical user interface to design and visualise arbitrary optical elements, set up wave propagation simulations and visualise their results. To ensure an efficient visualisation of the results, all simulation data are stored in a structured database that can filter and sort the output. Finally, we present a practical use case for Juno, where optical design and fabrication are interlaced in a feedback cycle. The presented data show how to compare the simulated and the measured propagation; if a difference or unexpected behaviour is found, we show how to convert and import the optical element profile from a profilometer measurement. The propagation through the profile can provide immediate feedback about the quality of the element and a measure of the effects brought by differences between the idealised and the actual profile, therefore, allowing to exclude the experimental errors and to weigh every aspect of fabrication errors.

## 1. INTRODUCTION

Accurate simulation and analysis tools ensure fast and efficient design, making research and engineering developments more efficient. These numerical tools verify new ideas and provide the basis for rapid prototyping by allowing the system's performance. This analysis would be costly and time-consuming, especially when increasing the system size and complexity.

In optical system design, there are multiple methods to approximate light propagation with various levels of precision, such as the relatively computationally efficient ray tracing [1], or more intensive grid-based differential numerical modelling methods [FTDT] [2], Fourier-based propagation [3] and many others. Depending on the elements used and the distances involved, one or the other is preferred. In ray tracing, light is treated as if it were travelling in straight lines that deviate upon interaction with a surface; the deviation is derived from Snell's Law, accounting for the angle of incidence and the difference between the refractive indices of the media [4]. When dealing with simple optical elements, such as lenses, colour filters and apertures, ray optics is sufficient, allowing for a simple framework to optimise the positions and profiles of lenses. Multiple simulation packages such as Zemax [1], Soltrace [5] and TracePro [6] have built-in optimisation procedures.

However, the behaviour of complex, multi-element optical systems often cannot be explained sufficiently by ray optics; diffractive elements, such as gratings or filters, require numerical modelling of wave propagation [7-9]. Here, any parametric optimisation, such as those used in ray optics, often cannot be easily performed. Therefore, designing new optical systems requires a more exploratory approach involving extensive screening to define the optimal parameters. An easy-to-build flexible simulation environment is greatly beneficial to minimise the time from design conception to parameters scan to evaluate the idea.

This type of exploratory simulation is directly applicable to investigations of new methods for beam shaping. Beam shaping has many uses in optics; for example, over the past decade, light sheet imaging and structured illumination fluorescence microscopy provided optical sectioning, reduced photodamage and fast super-resolution imaging. Other uses include optical trapping, where the position of dielectric particles is controlled through electric field gradients. Modifying light propagation in new ways for these purposes inherently requires novel optical elements that do not have well-studied behaviours. These beam-shaping applications also often rely on diffractive elements, necessitating wave propagation simulation. While simulation packages to model diffraction effects exist [10-12], there are several issues when attempting to use these packages for exploratory light propagation simulations.

Commercial wave propagation simulations typically come with a high cost, have a steep learning curve, and are not flexible enough to integrate custom analysis and novel optical components. Commercial options are also often made with a focus on optimisation, beginning with a known desired system behaviour and estimated element placement; numerical optimisation is carried out on element profiles and positions to maximise performance. Open-source options are often application-specific. Typically, the options designed for general use are, for the most part, scripted, requiring users to know the programming language used and, most importantly, to build the simulation structure themselves. Despite the wide adoption of Python over the past few years, the overhead of designing an entire optical system without an interactive visual output can be overwhelming when first approaching this topic and can result in a considerable threshold for implementing a functional package.

There are cases where standard element libraries do not include the desired optical elements. For example, lenses whose surface is defined by high-order polynomials as those used for aberration correction need to be manually defined. Here, in an exploratory study, the modification of the element parameters may require custom scripting, particularly when sweeping through a broad parameter space. This means that even defining the simulation is an involved procedure requiring great care from the user. Finally, examining the results is time-consuming and often complex without a structured data framework based on rigid parameter tracking and integrated visualisation tools.

To ease the process of designing new optical systems benefiting from wave propagation, we introduce the Python-based package Juno: an easy-to-use and modular package built to standardise exploratory light propagation simulations.

## 2. THE JUNO PACKAGE

Juno is a set of modular Python interfaces and functions enabling the accessible design of optical systems. The propagation functions are standardised to accept commonly used NumPy-style arrays [], which allow for a combination of custom optical elements or media in sequence. This feature enables users to build a wide variety of possible simulation pipelines while providing a standardised and intuitive framework that allows quick feedback cycles detailing the system's performance.

All simulated systems can be defined by parameters that can be tracked and used to filter numerical results for visualisation. The desired system parameters are stored in YAML configuration files [13], which are parsed to the visualisation module, allowing users to quickly inspect the effects of changing individual parameters for further performance optimisation. The graphical user interface of Juno was built as a plug-in for Napari [14], which ensures portability and allows its extension thanks to the built-in visualisation.

Juno is distributed as an open-source Python package through the MIT Licence, with documentation available on GitHub through https://github.com/DeMarcoLab/juno.

**Graphical User Interface**

The graphical user interface provides a systematic and intuitive method for creating, running and visualising simulations. It consists of five main windows, with an opening menu to enter each window individually (figure 1).

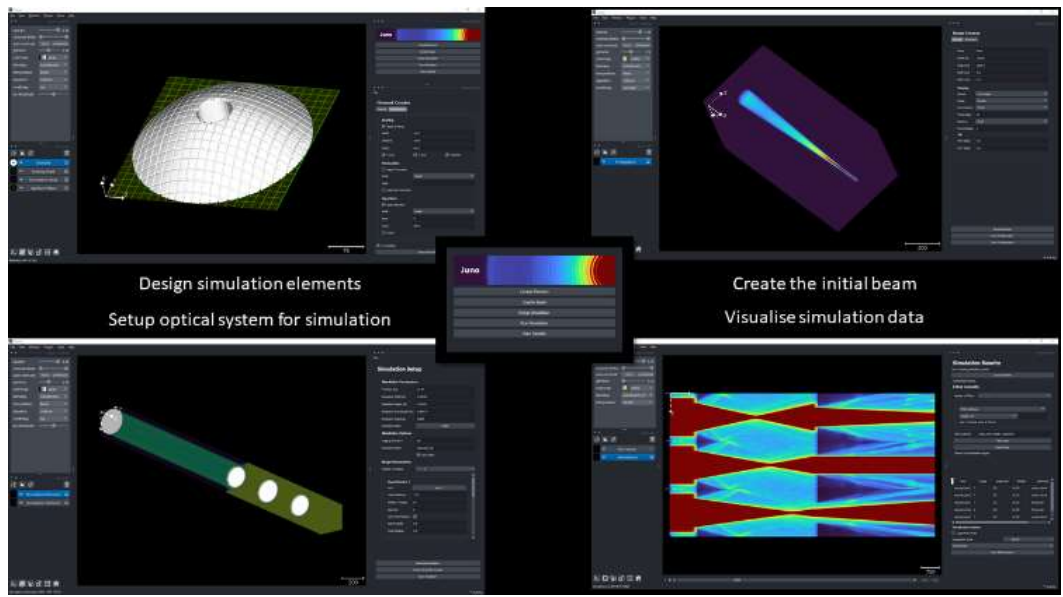

Fig. 1. The launcher interface and the structure of the Juno modules. The panel in the centre of the figure can call all the individual steps displayed around.

**Propagation method**

In Juno, the wavefront propagation is implemented under the thin lens approximation. The optical elements are modelled as thin phase-plates that introduce phase-shift and absorption onto the propagating wavefront. The propagated wavefront to a distance upstream is estimated in the Fresnel approximation by calculating the Fresnel integral using the Fourier Transform method [3]. This is a balanced option between simulation accuracy and speed. Due to the modular nature of Juno, it is simple to insert any propagation method desired whilst retaining the benefits of the package.

In Juno, the propagation is standardised as a series of "field and propagation" blocks that are sequentially stacked. The first step is the definition of the beam properties, which can be achieved using the BeamCreator module. Then typically, an optical element is defined at the beginning of the first block. Optical elements can be any phase plate with which the incident wavefront interacts. This produces the 'field' component, a 2D array describing the electric field distribution. Each field is then propagated until the next element in the system. Each optical element can be either generated through the ElementCreator, or by specifying the function describing the surface topology through custom scripts. These elements can then be inserted into the block (figure 2).

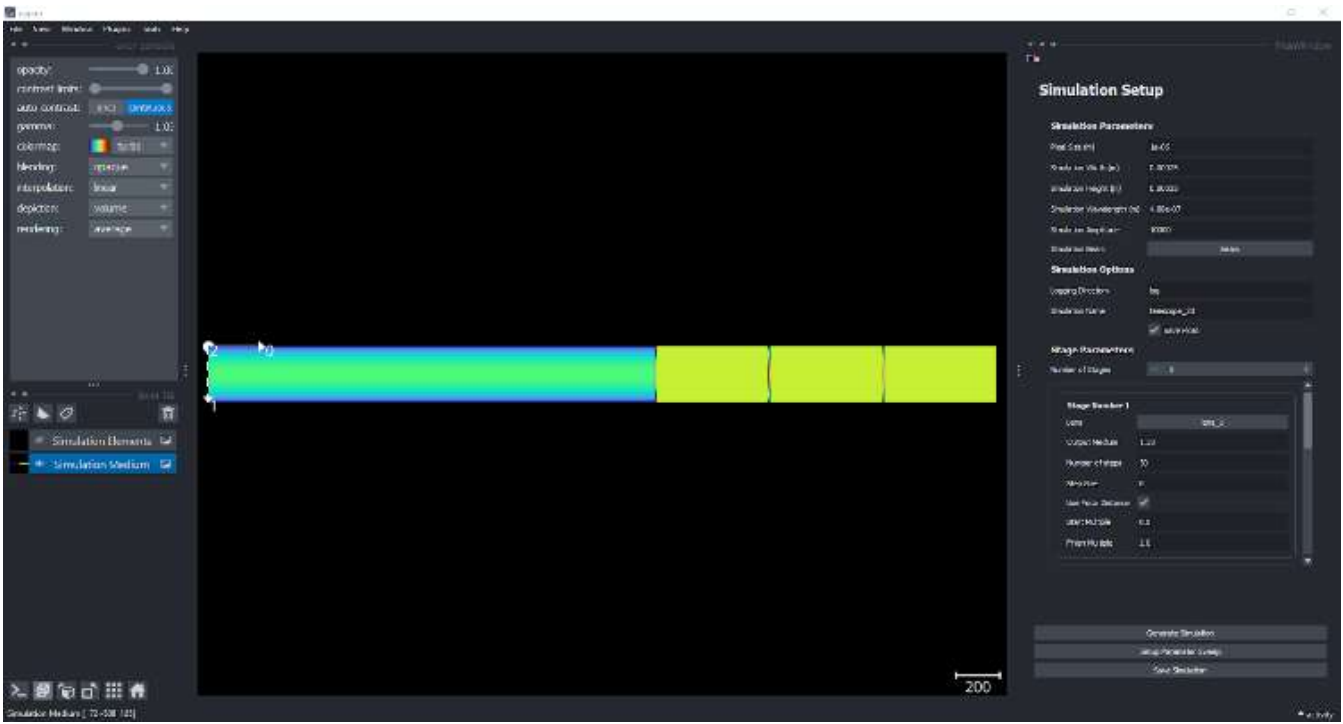

Fig. 2. The setup of the simulation blocks.

Since the propagation to any distance does not depend on the wavefront at any other plane, it is possible to reduce the computational time required for extensive simulations if only specific regions of the propagation are of interest. For example, in the case of a telescope, the propagation can be simulated as two planes, which requires only one computational step before the field is propagated out of the second plane (corresponding to the second lens).

As mentioned above, the propagation is performed under the thin lens approximation; therefore, to describe an optical element with a discrete thickness, the most straightforward approach consists in propagating the wavefront through two distinct surfaces defining the element boundary with the refractive index of the medium between the two surfaces defined by the material of the element.

## Beam Creation

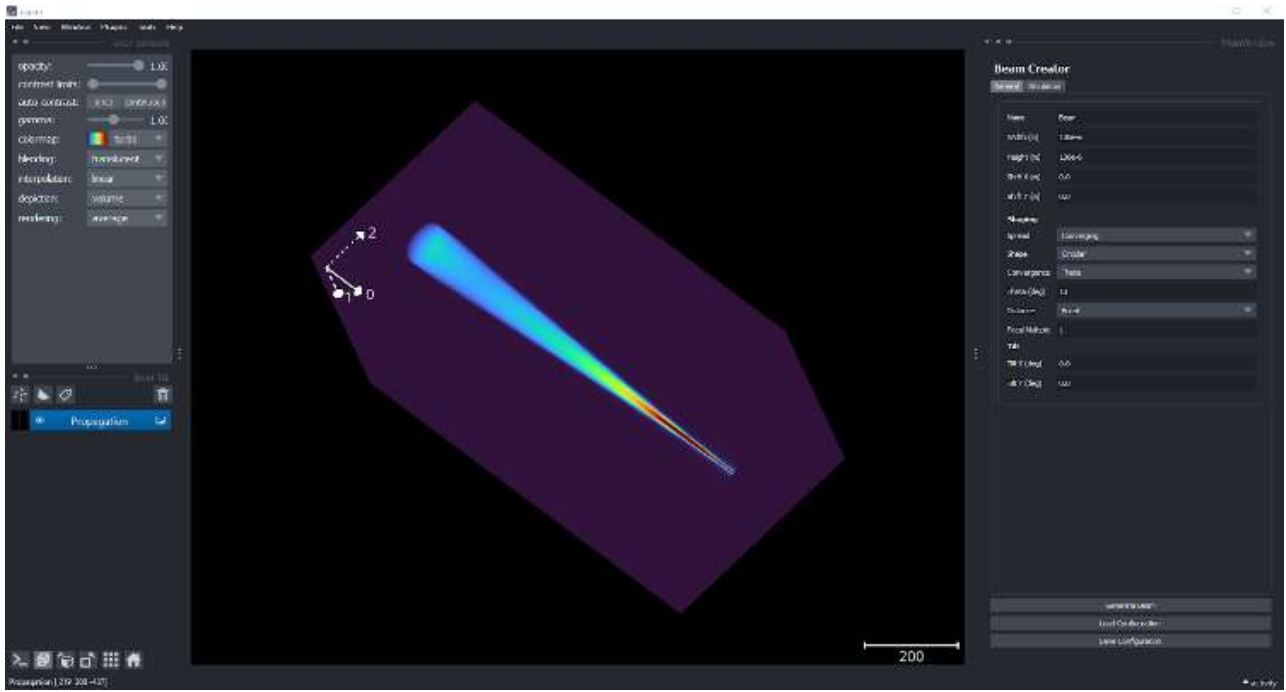

Fig. 3. The Beam Creation interface. An example of initial beam propagation through the interactive 3D visualisation. The interface will automatically render the propagation of the beam in 3D and show the beam incident to the first optical element in the system.

### *Beam Properties*
Beams are created with a minimal number of parameters required which are all visible in figure 3 (e.g. the wavelength and the size of the wavefront at the source). The user interface supports the creation of beams with different spreads (Planar, Converging and Diverging) and shapes (Circular and Rectangular). The user specifies the desired properties (e.g. final beam width or half-angle), and the package will calculate the corresponding wavefront. The supported properties include Beam Tilt, Beam Spread, Beam Shape, and Propagation Distance. Typically, the beam is first created, and in a second moment, tilt and shift are applied.

The interface will automatically update and restrict the user from creating non-valid configurations. Beam configurations can be saved and loaded in multiple setups.

# Definition and insertion of optical elements

The Element Creation interface provides the means to quickly generate, modify and visualise the elements used in the simulation. It consists of two tabs, General and Modifications. Elements, once defined, are saved to a configuration file that can be loaded in Juno at any time (as it happens for the Beam configurations).

Optical elements are generated and modified live; the newly designed components can be inspected interactively within the Element Creator as variations are applied (figure 4).

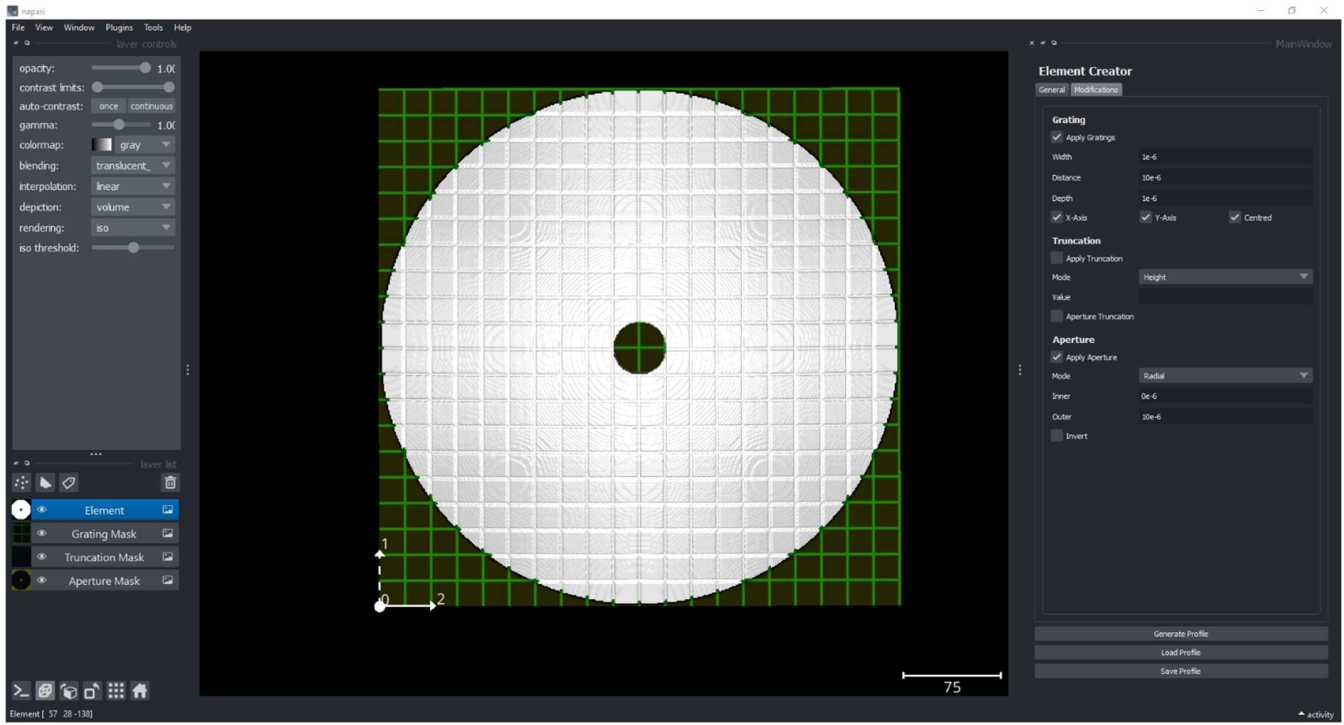

Fig 4. The Element creator interface. Through this interface it is possible to easily design arbitrary elements and modify them to with apertures and diffraction gratings.

### *General tab*

The general element properties allow the creation of cylindrical and spherically symmetrical lenses. Base lenses can also be loaded from a pre-existing configuration file or numpy array. This enables the viewing and modification of custom lenses and adjustments to previously generated lenses created in Juno.

### *Modifications tab*

Juno supports three types of modifications natively:

(i) Diffraction gratings can be applied to any surface as horizontal, vertical or in both directions. Typically, this results in phase gratings, although if applied to a non-transparent material, it will produce an absorption grating.

(ii) Apertures can be circular or square, with inner and outer radii. The aperture can also be inverted to allow light only through the aperture instead of blocking it.

(iii) Truncation is defined by a radius from the element's centre or by the absolute height to which one wants to truncate it. Truncations can also be applied as apertures.

## Simulation Setup

The Simulation Setup window is where the individual components assemble into an optical system. The optical elements and propagation mediums are both visualised as a single system. Each simulation is constructed from a set of global parameters, an initial beam, and multiple stages. The user can set the global simulation parameters and select the components previously created through the Element and Beam Creation interfaces (figure 5).

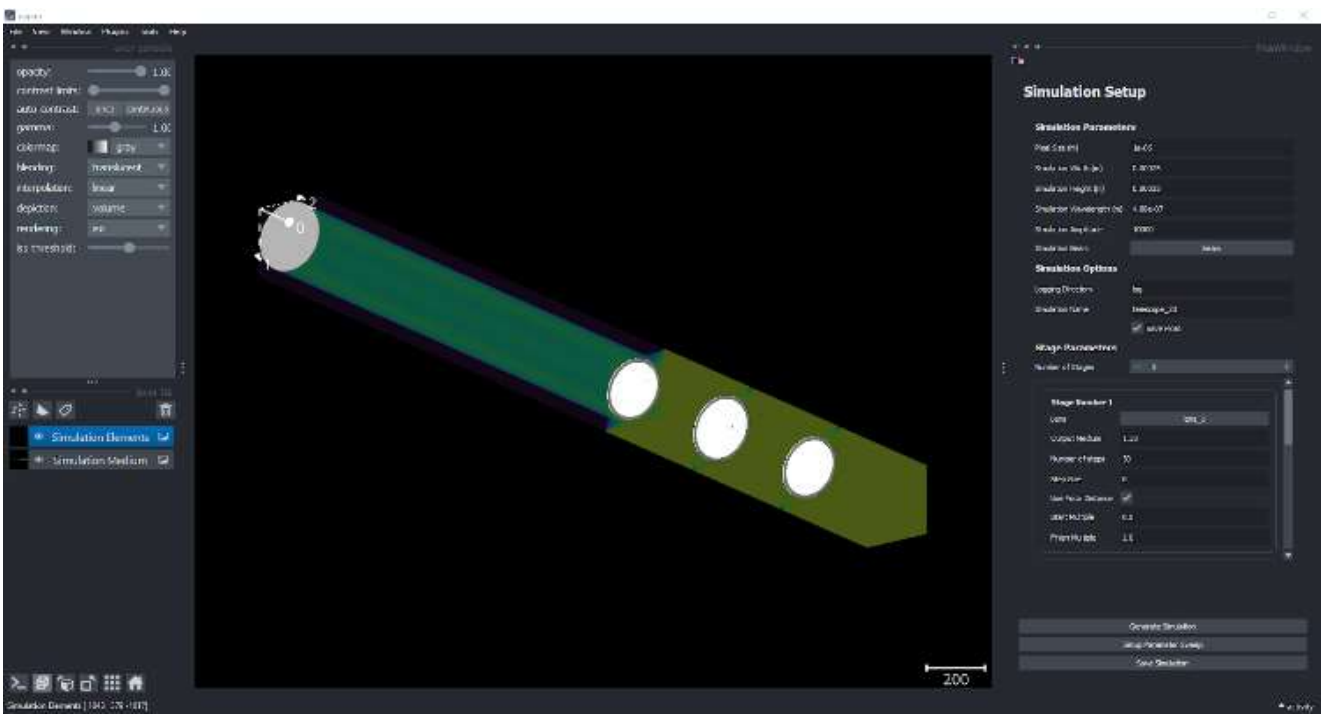

Fig 5. The Simulation Setup interface shows an example of a multi-stage system (Telescope).

The Simulation setup has two options: 'Sim Width' and 'Sim Height', designed to visualise how the beam propagates. Tuning of these parameters ensures that the beam does not exceed the simulation bounds. Once the components are selected, the package will validate the simulation and present a visualisation of the system to the user (Figure 4).

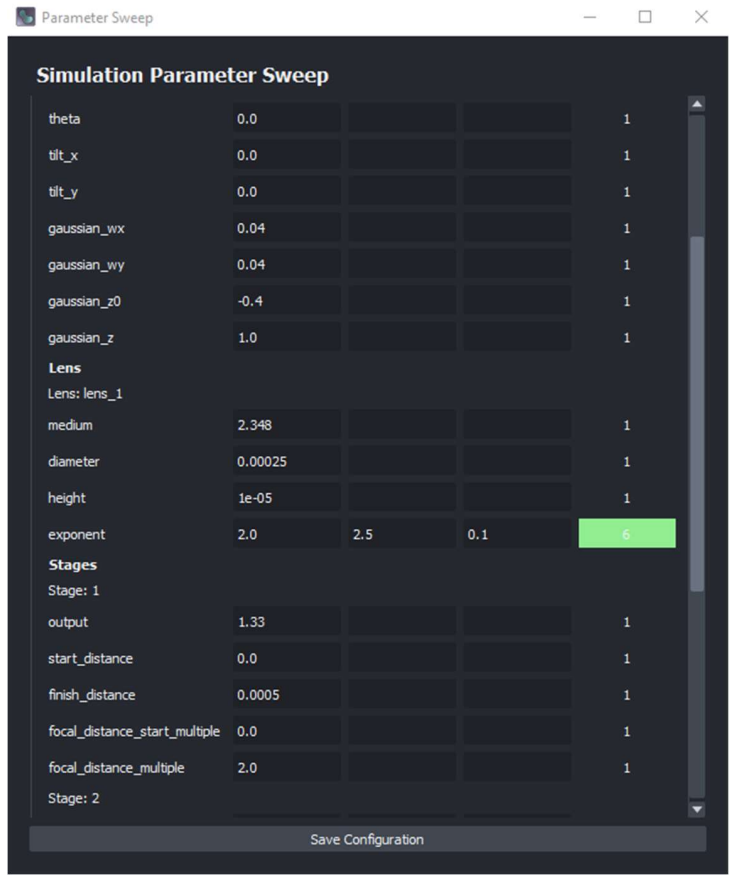

Fig 6. Juno's parameters sweep user interface. Independent ranges for all the parameters defining the optical elements can be introduced. These include position, curvature or function defining the surfaces, and refractive index of the media.

After the validation, it is possible to define a parameter sweep across all available numerical parameters (Figure 6), enabling rapid experimentation and iteration. All required simulation information is saved into a single YAML file, called the simulation configuration file.

## Simulation Runner

This panel guides the user through the actual simulation procedure. It allows loading and running simulations previously configured through the Simulation Setup interface. The package will validate the configuration and generate all the simulations (if a parameter sweep is specified).

When the simulation runs, a progress bar indicates the completion of all the simulations in the sweep. Each simulation in a sweep is given a randomly generated name to identify it. Once complete, metadata and results are available in the selected directory.

## Visualise Results

Depending on the complexity of the simulation and how extensive the parameter sweep has been, the results can be extremely complex to interpret. Accordingly, the design of this interface provides the means to load, filter and visualise all simulation results in a structured manner. As such, any available metadata can be used for filtering and sorting. For example, the elements parameters or any custom-defined metadata that has been manually associated with any of the elements.

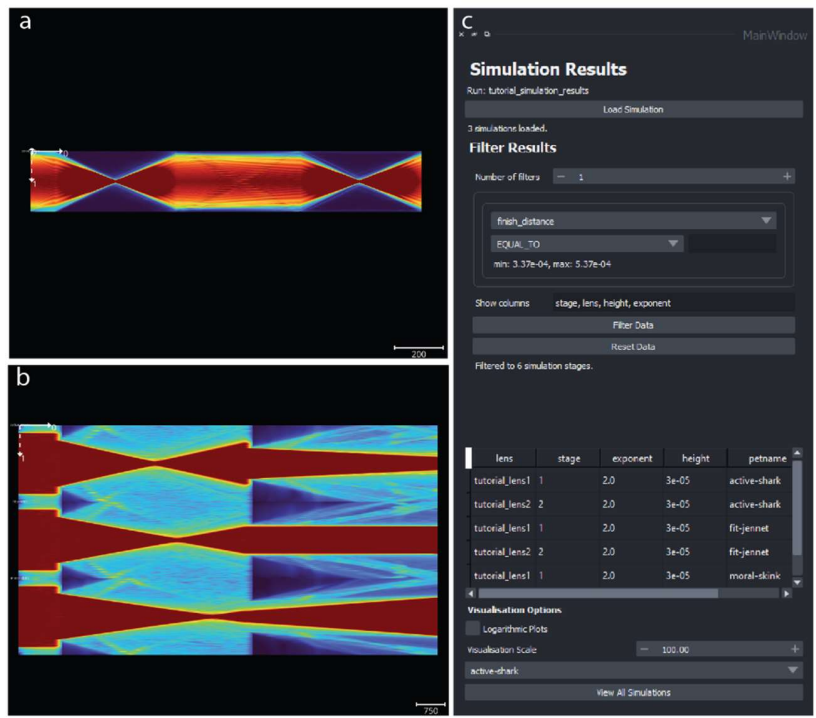

Fig 7. Example of visualisation of a single telescope simulations in (a) and three different telescope simulations in (b). Both outputs are displayed using a 2D ortho-slice view. This is a quick and efficient way to visualise the propagation through the optical path. (c) shows the selection panel allowing to choose which simulations to display and at what scale.

All the simulation data is displayed in the centre view (Figure 7). Users can cycle through different views using the built-in controls to see slices along any simulation axis. Simulation data can also be viewed as an interactive 3D volume (Figure 8); however, this rendering is heavier on the graphic hardware.

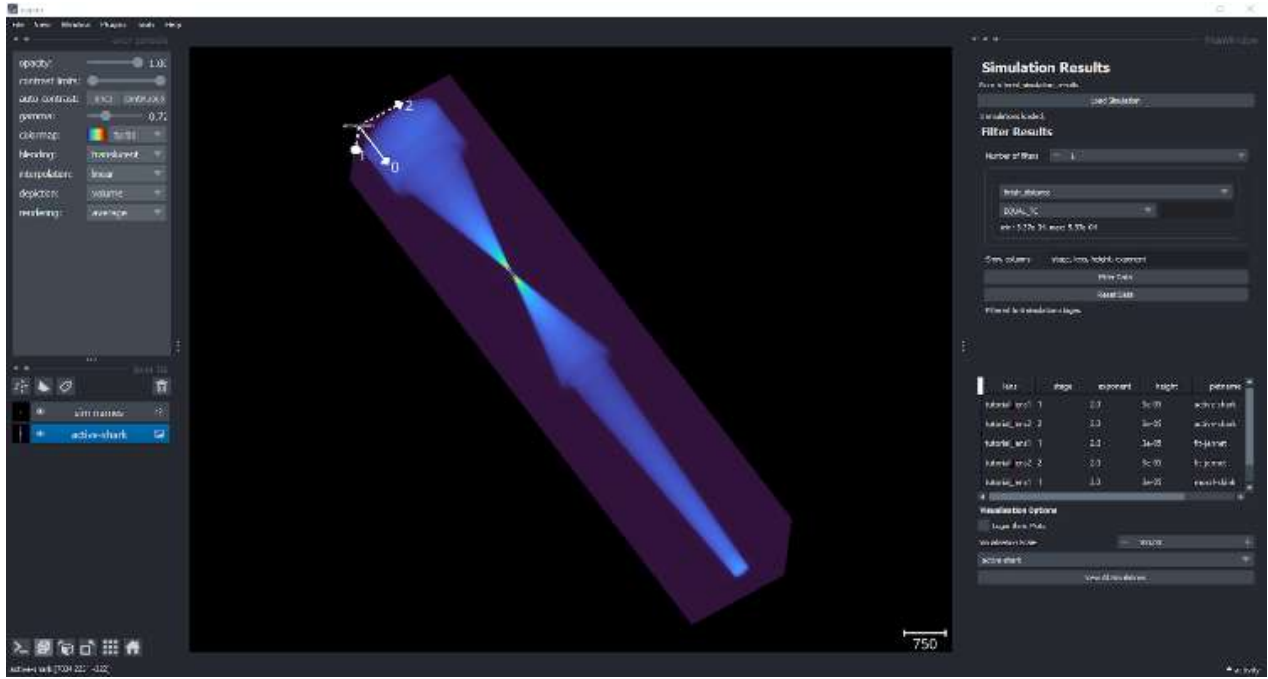

Fig 8. 3D interactive visualisation of a single simulation output. The three-dimensional view provides the means for understanding the propagation of complex or asymmetric beams.

## 3. Example use case for Juno to improve optical design and fabrication

Here we present an exploration of optical element design, simulation and characterisation using Juno. This example analysis builds upon an earlier study conducted by our lab on the development of hybrid diffractive refractive elements [15], utilising the new simulation package to improve the process.

Diffractive gratings can be embedded in microlenses to create hybrid elements with both refractive and diffractive properties [16]. In this study, we embed diffractive gratings into axicons in both x and y directions, multiplexing Bessel-like beams into a diverging two-dimensional lattice. The elements are fabricated from lithium niobate using the method described in Gorelick et al. [17]. The simulations and output measurement were conducted with a wavelength of 458 nm, from 0-2 mm and propagated through air.

Using the Juno interface, an axicon with a height of 7.5 µm, diameter of 250 µm and grating depth of 300 nm was generated, and its output simulated. In initial simulations, most energy is contained in the 0th order (Figure 3a, d). After fabrication, a 20x, 0.45 NA objective (Olympus MPlanFL N) was used to measure the propagation (Figure 9 b, e). As can be seen, the fabricated lens diverges energy into a range of diffraction orders.

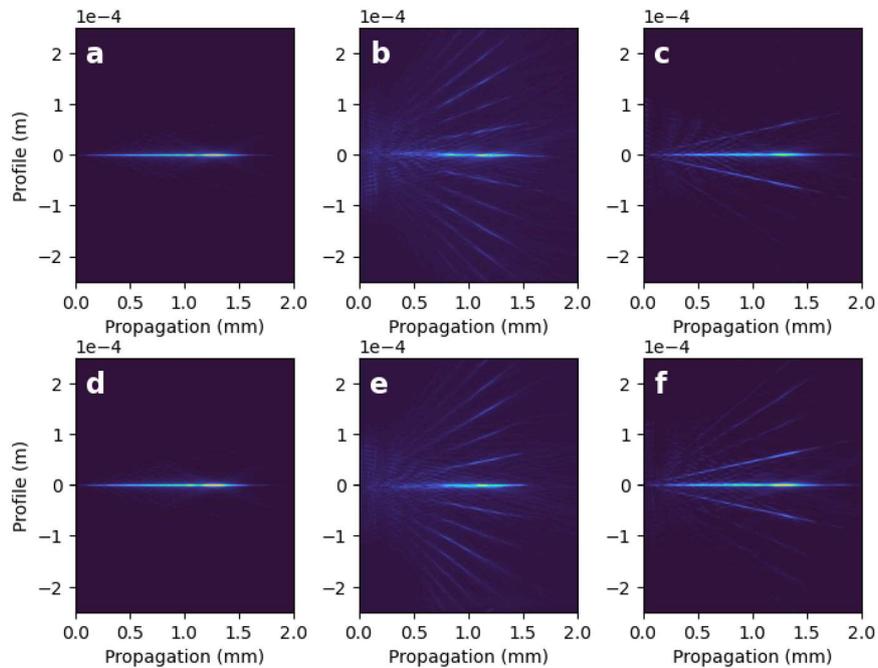

Figure 9. Intensity profiles from top-down (top row) and side-on (bottom row) from 0 - 2 mm: (a, d) output from an ideal simulated refractive, diffractive axicon. (b, e) output measured from the fabricated element. (c, f) output from the measured surface profile showing good agreement with a fabricated axicon.

To explain the difference in performance, white light interferometry was used to obtain the surface profile of the fabricated element (Figure 10a). Upon examining the profile, the measurements showed sloped grating walls due to the resolution of the manufacturing process, compared to perfectly squared simulation gratings. To characterise the differences and better understand the effect that manufacturing limitations can have on the desired outcome, we converted the interferometry measurements into a Juno-compatible profile (Numpy array). We imported it into Juno to simulate the propagation through the measured profile. Given the gratings' depth, there have been regions where the measurements produced non-numerical data (Figure 11); these were corrected through local interpolation.

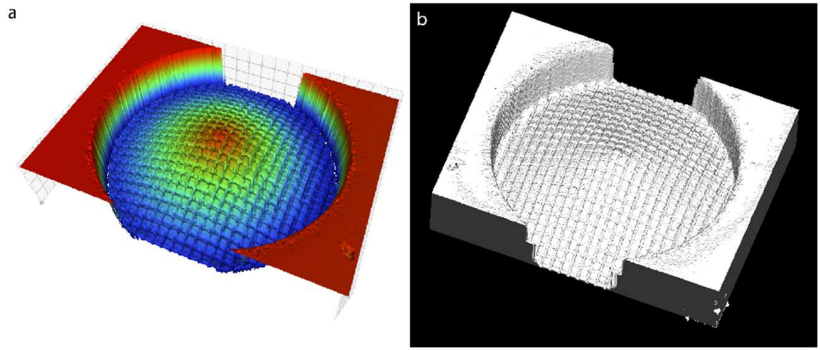

Figure 10. Three-dimensional surface profiles of a hybrid diffractive-refractive element with 250 μm diameter, 7.5 μm height and 300 nm grating depth in x and y: (a) white light interferometry measurement of the surface profile. (b) Recreation of the surface profile in Juno's ElementCreator interface

The possibility of simulating the propagation through the actual profile increases the confidence in the measurements and accounts for any additional divergence introduced during the manufacturing process. The example above shows the effects of limited manufacturing resolution on the elements' performance. There are many other unwanted collaterals during manufacturing that may be limiting or changing the performance of the elements; for example, material redeposition would not be accounted by an idealised profile, but its presence induces non-specific scattering; alternatively, differences in the local curvature due to the geometry will change the effective performance. The methodology can also be applied to any other surface profile without requiring additional measurements of the profile slopes. The new arrays were inserted directly into the user interface (Figure 10 b), with the simulation pixel size matching the profilometers' resolution.

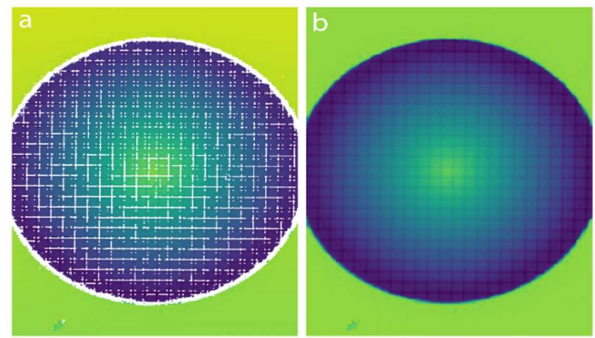

Figure 11. Processing of the element surface profile. (a) missing data from the measured interferometry profile (white) showing areas of sharp depth change. (b) processed profile used in the second set of simulations.

Good agreement between the measured light intensities and updated simulations (Figure 9 c, f and Figure 12) show that fabrication limitations cause most deviations from the initial simulations. Future element design shall account for the sloping in the fabrication process. This isolates issues beyond the simulation, as any extra discrepancies in measured results can be attributed to experimental factors.

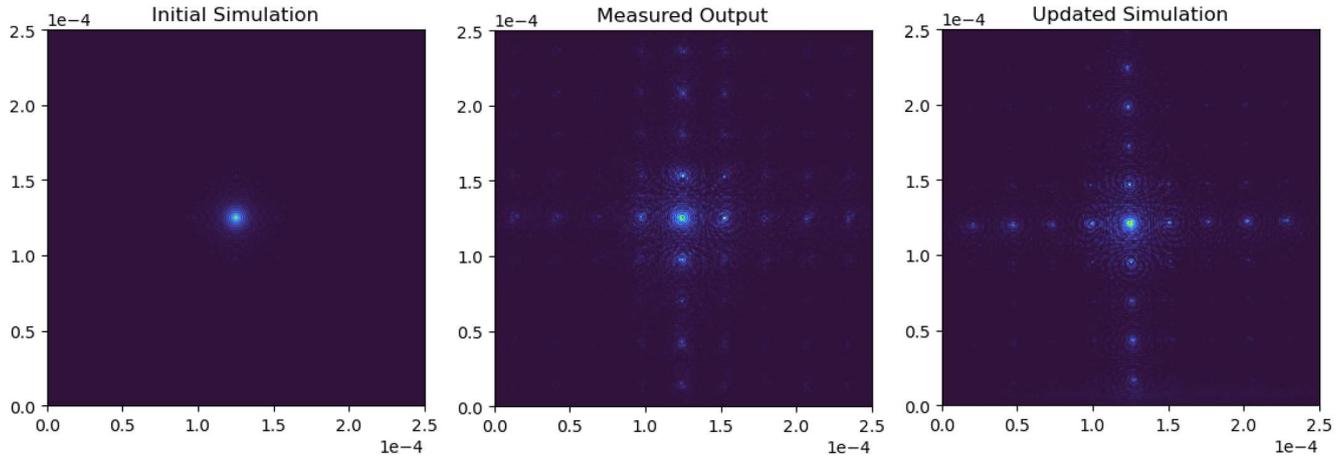

Figure 12. Intensity maps transverse to the optical axis at 1.2 mm along propagation: (a) idealised element profile generated in Juno. (b) The measured output from fabricated elements. (c) The simulated output of measured surface profile element.

## 4. SYSTEM REQUIREMENTS

Juno has been developed using the scientific Python packages numpy [18], scipy [19], pandas [20], dask [21] and zarr [22]. The control is done through configuration files using pyyaml [13] in conjunction with petname ✓ used to keep track of simulation results. The graphical user interface is built using PyQt5 and Napari [14]. Also utilised are tqdm [23], matplotlib and imageio [24]. Finally, the pytest [25] suite is used for testing.

### *Running simulations on HPC*

As the simulation is generated from a single configuration file, it is possible to run the simulation on any system. Accordingly, one can create a simulation configuration file locally using the Juno interface and transfer it to a high-performance computing system with the appropriate Python environment.

### *Benchmark*

To benchmark the performance of the simulation package, we ran the tutorial simulation on a standard laptop (16GB RAM, 11th Gen Intel(R) Core(TM) i5-1135G7 @ 2.40GHz 1.38 GHz CPU). The simulation consisted of two lens stages and used a pixel size of 1um, with a 400 μm × 400 μm simulation size and 110 simulation steps. Running this simulation with the full suite of data saving and visualisation took 4 minutes and 11 seconds.

## 5. LIMITATIONS

Juno does not currently use the speed-ups that come from distributed computing or computation through GPUs, as in our hands, the size of the simulations did not yet require those implementations. Given the modular nature of the package, implementing GPU or multithread computations will be easily possible. We plan to continue developing the package to include these features and support new use cases.

Juno is currently focused on wave propagation and does not use ray optics. The reason for this choice lies in the inaccuracy of ray propagation in the case of diffractive optics and the lack of precise edge effects on the apertures' edges. Accordingly, we do not plan the development of the package in this direction.

The simulations' analysis currently consists of the visual inspection of the output through a structured selection based on the associated metadata. The visualisation is interactive and offers 2D and 3D representation. Since all the output is standardised to be compatible with Numpy arrays, it is simple to add custom analytical pipelines. Any analysis pipeline we will develop will be added to the GitHub repository and will be made publicly available.

# 6. DECLARATIONS

## A. Author's Contributions

DD designed the program and developed the algorithms. PC developed the package architecture and user interface. SG developed the propagation. JCW and AdM conceived the project. DD and PC contributed equally to this work. All authors contributed sufficiently to the work to take public responsibility for appropriate portions of the content. All authors contributed to the manuscript and approved the final version.

## B. Funding

DD and JCW were supported by the Australian Research Council FL180100019 Laureate Fellowship program. PC was supported by the Australian Research Data Commons project XN006.

## C. Data availability

All the code is open source and freely available through GitHub at www.github.com/demarcolab/juno. The package is licensed through the standard MIT license.